\documentclass[runningheads]{llncs}
\usepackage{amsmath}
\usepackage{pst-node}% http://ctan.org/pkg/pst-node
\usepackage{multido}% http://ctan.org/pkg/multido
\usepackage{enumitem,xcolor}
\usepackage{tikz}
\usepackage{graphicx}
\usepackage{caption}
\usepackage{subfigure}
\usepackage[utf8]{inputenc}
\usepackage{pgfplots} 
\usepackage{pgfgantt}
\usepackage{pdflscape}
\usetikzlibrary{spy}
\usetikzlibrary{positioning,calc}
\usetikzlibrary{decorations.pathmorphing,calc,shapes,shapes.geometric,patterns}
\usetikzlibrary{shapes.multipart}
\usepackage{xfrac}
\usepackage{colortbl}
\usepackage{amssymb}
\usepackage{cancel}
\usetikzlibrary{arrows,positioning,calc,intersections}
\usetikzlibrary{datavisualization.formats.functions}

\newcommand{\setx}{\ensuremath{\mathcal{X}}}
\newcommand{\sety}{\ensuremath{\mathcal{Y}}}

\title{Joint Detection and Identification for Scalable Control of Nanorobot Swarms under Harsh Communication Constraints}

\titlerunning{Joint Detection and Identification}

\author{
Wafa Labidi\inst{2} \and
Holger Boche\inst{2,4} \and
Christian Deppe\inst{1} \and
Marc Geitz\inst{3}
}

\institute{
Technische Universität Braunschweig, Germany \and
Technical University of Munich, Germany \and
T-Labs, Deutsche Telekom \and
Munich Center for Quantum Science and Technology
}

\begin{document}

\maketitle
\begin{abstract}
The coordination of large populations of highly constrained devices, such as micro- and nanoscale agents in biomedical applications, poses fundamental challenges to classical communication paradigms. In scenarios such as targeted drug delivery, devices operate under severe limitations in energy, size, and communication capabilities, while requiring precise and selective activation within spatially localized regions.

In this work, we propose the framework of Joint Detection and Identification (JDAI) as a system-level approach for scalable control under such constraints. The key idea is to shift from reliable message transmission to a control-oriented paradigm, in which devices locally decide whether a broadcast signal is relevant. This enables implicit addressing and subset activation without the need for explicit per-device communication.

We demonstrate how message identification can be combined with sensing. This enables the realization of a closed-loop system that integrates detection, communication, and actuation. Using the example of targeted nanorobot therapy, we analyze the interplay between sensing resolution, communication constraints, and system dynamics. In particular, we show that while identification exhibits favorable asymptotic scaling, practical implementations are governed by finite blocklength effects, noise, and latency.

The proposed framework complements existing physical-layer communication approaches, including molecular, electromagnetic, and acoustic techniques, by providing a control-layer abstraction for scalable subset selection. Overall, JDAI connects identification-theoretic principles with system-level design to control large, resource-limited environments.
\end{abstract}

\section{Introduction}

Emerging application domains such as large-scale Internet-of-Things (IoT) systems, distributed autonomous agents, and in-body micro- and nanorobotics require the coordination of massive populations of highly constrained devices \cite{akyildiz2008nanonetworks,akyildiz2010internet}. In many such scenarios, individual devices are severely limited in terms of energy, computational capabilities, size, and communication bandwidth. As a consequence, classical communication paradigms based on explicit addressing and reliable message transmission become increasingly inefficient, or even infeasible, as the number of devices grows.

This challenge is particularly pronounced in biomedical applications such as targeted drug delivery using micro- and nanoscale agents \cite{martel2012magnetic,sitti2009miniature}. In such environments, devices are often passively transported, for example through the bloodstream, and operate under strict constraints on size and power consumption. Communication between an external control system and the devices may be limited to low-rate broadcast mechanisms, for instance based on magnetic, acoustic, biochemical, or other highly constrained signaling modalities \cite{farsad2016survey,nakano2013molecular,akyildiz2014internet,guo2018ultrasonic,sun2010magnetic}. At the same time, precise control over subsets of devices is required, e.g., to activate therapeutic agents only within a spatially localized region such as a tumor site. This raises a fundamental question: how can a control system selectively activate relevant subsets of devices without explicitly addressing each device individually?

In many relevant applications, it is not necessary that every individual device reacts reliably. Instead, it is often sufficient that a subset of devices located within a region of interest performs the desired action. The system now focuses on reliably controlling a subset of devices, not on delivering high-rate data to each one. This observation shifts the system objective from reliable per-device communication to robust subset activation. Hence, the problem is not primarily one of high-rate data delivery, but of scalable control under extreme communication constraints.

A broad range of communication paradigms has been proposed for nanoscale and in-body systems, including molecular communication \cite{farsad2016survey,nakano2013molecular}, terahertz electromagnetic communication \cite{akyildiz2014internet}, ultrasound-based links \cite{guo2018ultrasonic}, and near-field magnetic induction approaches \cite{sun2010magnetic}. These paradigms differ significantly in terms of achievable rates, propagation characteristics, and implementation complexity. However, many of them focus primarily on physical-layer data transmission and networking, while the problem of scalable control and selective activation under extreme communication constraints remains less explored.

In this work, we show that message identification offers a new approach that is particularly well suited for such control problems. Identification via channels, introduced by Ahlswede and Dueck \cite{ahlswede1989identification}, departs from the classical transmission paradigm by focusing on the problem of deciding whether a specific message was sent, rather than reconstructing it. A key result is that, under randomized encoding, the number of identification messages (also called identities) grows double-exponentially with the blocklength, in contrast to the exponential growth in classical transmission \cite{ahlswede1989identification,han2003information}. While this scaling law is asymptotic and does not directly translate into finite-blocklength performance, it indicates a fundamentally different operating regime that is highly attractive for large-scale control systems.

From a system perspective, identification enables receivers to locally decide whether a broadcast signal is relevant. This naturally leads to a form of implicit addressing, where subsets of devices can be selected without requiring the communication to each device one by one. In the scenarios considered here, an ``identity'', that is an identification message, should therefore not be interpreted as the label of a single physical device, but rather as a control instruction, for example specifying a spatial region, a temporal condition, and a desired action. Each device evaluates the received signal based on locally available information and performs a binary decision on whether the instruction is relevant. In this sense, identification provides a natural control-oriented abstraction for subset activation in large populations of simple agents.

An additional important aspect is the role of shared randomness and feedback. It is known that common randomness can enhance identification performance additively, and such randomness may in principle be generated dynamically through interaction, sensing, or feedback mechanisms \cite{ahlswede2008general,ezzine2020common,wiese2022identification,labidi2025joint}. This observation further strengthens the relevance of message identification in settings where sensing and communication are tightly coupled.

Building on these principles, we propose the concept of \emph{Joint Detection and Identification} (JDAI) as a system-level framework that integrates sensing and message identification. The key idea is to decouple the control process into two tightly coupled stages: a sensing stage that identifies candidate regions of interest, and a communication stage that enables devices to locally decide whether to react to a broadcast signal. In this architecture, the control system does not explicitly address individual devices; instead, it detects where intervention is needed and then triggers appropriate actions in the relevant subset of the population through identification-based signaling.

To illustrate this concept, we consider a biomedical application scenario in which a large population of micro- or nanoscale devices is deployed for targeted drug delivery in the human body. In such a setting, sensing mechanisms such as magnetic resonance imaging (MRI) or related techniques provide coarse spatial information at a resolution that depends strongly on the sensing modality, acquisition time, and physical contrast mechanisms \cite{gleich2005tomographic}. At the same time, communication may be limited to extremely low-rate signals, for example based on low-frequency magnetic field modulation. Under these constraints, the control problem shifts from reliable message transmission to scalable subset selection.

It is important to emphasize that message identification does not replace existing physical-layer communication paradigms. Rather, it provides a complementary abstraction at the system level. In particular, the JDAI framework can be interpreted as a control-layer mechanism that operates alongside different physical-layer technologies, including molecular, electromagnetic, acoustic, or magnetic communication approaches.

In this context, molecular communication (MC) has been identified as a key enabler for communication among nanoscale devices in biomedical environments. 
For an integration of MC into future 6G systems, we refer to \cite{Hasel2019}. A comprehensive overview of MC technologies and their potential for 6G applications is provided in \cite{Bien2026,Schwen2023}. 
In contrast to classical communication paradigms, MC systems are inherently constrained in terms of data rate, latency, and reliability. 
As a result, communication in such systems is often goal-oriented and event-driven rather than based on the transmission of detailed messages. 
The post-Shannon communication paradigm considered in this paper is therefore well aligned with the requirements of MC systems. 
In particular, identification-based communication provides a natural abstraction for MC scenarios, where devices only need to decide whether a received signal is relevant and whether to trigger a specific action. 
This is consistent with the envisioned integration of MC into future 6G systems, where efficient, low-complexity, and task-oriented communication mechanisms are required.

The goal of this paper is not to provide a complete physical realization of such a system, but to establish a conceptual and system-level framework that connects identification-theoretic principles with emerging application scenarios. More specifically, we demonstrate how identification can be interpreted as a mechanism for selective activation, how it can be integrated with sensing, and how it enables scalable control in large populations of autonomous devices. Furthermore, we discuss key limitations arising from finite blocklength, noise, latency, and physical-layer constraints that must be addressed in practical implementations. Recent advances in structured and practical identification coding further indicate that such ideas are not only of theoretical interest, but may also serve as useful building blocks in realistic systems \cite{ferrara2022implementation,vonlengerke2023digital,vonlengerke2025tutorial,perotti2024wakeup}.

The remainder of this paper is organized as follows. Section~2 reviews the fundamentals of identification via channels and highlights the key properties for scalable control of large device populations. Section~3 introduces the JDAI framework and its operational principles. Section~4 presents a biomedical application scenario and illustrates the corresponding system architecture. Section~5 discusses feasibility aspects and system-level considerations. Finally, Section~6 concludes the paper.

\section{Fundamentals of Identification and Channel Model}

In this section, we review the concept of identification via channels and its fundamental differences from classical transmission. The goal is to introduce the key definitions and results underlying the JDAI framework, while maintaining a clear system-oriented interpretation.

\subsection{From Transmission to Identification}

The concept of identification via channels was introduced by Ahlswede and Dueck \cite{ahlswede1989identification} and represents a fundamental departure from Shannon’s classical communication paradigm \cite{shannon1948mathematical}.

In the classical transmission setting, a sender encodes a message $u \in \mathcal{M}$ into a channel input sequence, and the receiver aims to reliably reconstruct $u$ from the channel output. The objective is therefore accurate message recovery.

In contrast, identification considers a different task. The encoder selects an identity $v \in \mathcal{N}$, while the decoder is not interested in reconstructing $v$ itself. Instead, for a given query $v' \in \mathcal{N}$, the decoder only decides whether $v = v'$ or not. Thus, identification can be interpreted as a family of binary hypothesis tests, one for each possible identity.

A crucial aspect is that the encoder does not know the query $v'$ at the receiver. Therefore, the encoding must enable reliable decisions for all possible queries based on a single channel output.

\begin{figure}[t]
\centering
\tikzstyle{farbverlauf} = [ top color=white, bottom color=white!80!gray]
\tikzstyle{block} = [draw, fill=none, rectangle, rounded corners,
minimum height=4em, minimum width=1.5
cm]
\tikzstyle{block1} = [draw, fill=none, rectangle, rounded corners,
minimum height=4em, minimum width=4.5
cm]
\tikzstyle{blocked} = [draw, rectangle, rounded corners,
minimum height=2em, minimum width=2.5em,farbverlauf]
\tikzstyle{input} = [coordinate]
\usetikzlibrary{arrows}
\scalebox{.7}{\begin{tikzpicture}[auto, node distance=2cm,>=latex']
\node[](m){$v$};
   \node[blocked, right=1cm of m] (encoder) {Enc};
    \node[blocked,right=2cm of encoder] (channel) {noisy channel $W^n$ };
    \node[blocked, right=2cm of channel](decoder) {Dec};
    \node[ right=1cm of decoder] (mhat){$v^\prime$ transmitted or not?};
    \node[block,dashed, left=3.5cm of channel] (alice) {};
    \node[block1, dashed, right=3.5cm of channel] (bob) {};
    \node[above=.5cm of alice](a){Alice};
    \node[above=.5cm of bob](b) {Bob};
\draw[->,ultra thick] (encoder) -- node[above]{$x^n \in \setx^n$} (channel);
\draw[->,ultra thick] (channel) -- node[above]{$y^n \in \sety^n$} (decoder);
\draw[->,ultra thick] (m) --(encoder);
\draw[->,ultra thick] (decoder) -- (mhat);
\end{tikzpicture}}
\caption{Identification via channels. The encoder transmits an identity $v$ over a noisy channel, while the decoder evaluates a query $v'$ and decides whether $v = v'$. In contrast to classical transmission, the goal is not message reconstruction but a binary decision.}
\label{fig:ID_model}
\end{figure}
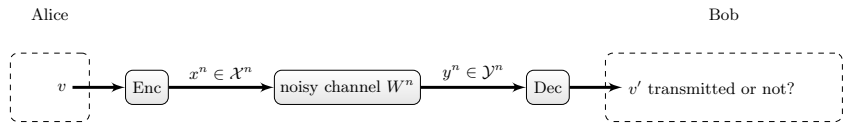

Figure~\ref{fig:ID_model} illustrates this fundamental difference. From a system perspective, this interpretation aligns naturally with control problems: a receiver evaluates whether a broadcast signal is relevant and decides locally whether to react. In the JDAI framework, identities correspond to control instructions rather than individual device labels.

\subsection{Channel Model}

We consider a discrete memoryless channel (DMC) defined by the triple $(\mathcal{X}, \mathcal{Y}, W)$, where $\mathcal{X}$ and $\mathcal{Y}$ denote the input and output alphabets and $W(y|x)$ is the transition probability \cite{cover2006elements}. For sequences $x^n \in \mathcal{X}^n$ and $y^n \in \mathcal{Y}^n$, the $n$-fold channel is given by
\begin{equation}
W^n(y^n|x^n) = \prod_{i=1}^n W(y_i|x_i).
\end{equation}

Random variables are denoted by upper-case letters and their realizations by lower-case letters.

\subsection{Transmission Codes}

A deterministic $(n, M, \lambda)$ transmission code consists of codewords $u_i \in \mathcal{X}^n$ and decoding sets $D_i \subset \mathcal{Y}^n$, and decoding errors that satisfy
\begin{align}
W^n(D_i^c | u_i) &\leq \lambda, \\
D_i \cap D_j &= \emptyset, 
\end{align}
for all $i,j=1,\ldots,M$ with $i\neq j$ and some $\lambda \in (0,1)$. 

Randomized transmission codes replace each codeword by a distribution over $\mathcal{X}^n$. However, randomization does not increase the achievable transmission rate.

\subsection{Identification Codes}

A deterministic $(n, N, \lambda_1, \lambda_2)$ identification code is given by codewords $u_i \in \mathcal{X}^n$ and decoding sets $D_i \subset \mathcal{Y}^n$, and with errors of the first and the second kind that satisfy 
\begin{align}
W^n(D_i^c | u_i) &\leq \lambda_1, \\
W^n(D_j | u_i) &\leq \lambda_2 \quad (i \neq j),
\end{align}
for all $i,j=1,\ldots, N$ with $i \neq j$ and some $\lambda_1+\lambda_2 < 1$.

In contrast to transmission codes, the decoding sets are not required to be disjoint.

A randomized identification code is given by probability distributions $Q(\cdot|i)$ over $\mathcal{X}^n$ and decoding sets $D_i \subset \mathcal{Y}^n$, and with errors of the first and the second kind that satisfy 
\begin{align}
\sum_{x^n} Q(x^n|i) W^n(D_i^c|x^n) &\leq \lambda_1, \\
\sum_{x^n} Q(x^n|j) W^n(D_i|x^n) &\leq \lambda_2,
\end{align}
for all $i,j=1,\ldots, N$ with $i \neq j$ and some $\lambda_1+\lambda_2 < 1$.

A key distinction from classical transmission is that randomization is essential to achieve optimal identification performance \cite{ahlswede1989identification,han2003information}.

\subsection{Capacity Results}
Let $W$ be a DMC. Let $M(n,\delta)$ be the maximal number $M \in \mathbb{N}$ such that an $(n,M,\delta)$ transmission code for $W$ exists. Let $N(n,\lambda)$ be the maximal number $N \in \mathbb{N}$ such that an identification code $(n,N,\lambda,\lambda)$ for $W$ exists. Let $C(W)$ be the Shannon capacity of $W$. Let $C_{ID}(W)$ be the identification capacity of $W$.

The classical channel coding theorem \cite{shannon1948mathematical} states that
\begin{equation}
C(W) = \lim_{n \to \infty} \frac{1}{n} \log M(n,\delta) = \max_{P_X} I(X;Y), \quad \text{ for }\delta \in (0,1).
\end{equation}

In contrast, the identification coding theorem \cite{ahlswede1989identification,han2003information} shows that
\begin{equation}
C_{\mathrm{ID}}(W) = \lim_{n \to \infty} \frac{1}{n} \log \log N(n,\lambda) = C(W), \quad \text{ for } \lambda \in (0,1/2).
\end{equation}

Thus, identification enables a double-exponential growth
\begin{equation}
N \approx 2^{2^{nC(W)}}.
\end{equation}

\subsection{Key Properties and System Implications}

The identification paradigm exhibits several distinctive properties particularly relevant for large-scale systems.

\textbf{Double-Exponential Scaling:}  
Identification allows selecting among extremely large sets of possible actions or device subsets using very limited communication resources. This behavior extends beyond DMCs to Poisson, Gaussian and MIMO channels \cite{labidi2021identification,EW2024}.

\textbf{Role of Common Randomness:}  
Common randomness between sender and receiver increases the identification capacity additively \cite{ahlswede2008general}. Such randomness can be generated via feedback or sensing mechanisms \cite{ezzine2020common,CR_TSPwithRami}.

\textbf{Feedback and Interaction:}  
Feedback can serve not only to improve reliability but also as a resource for generating shared randomness, thereby enhancing identification performance \cite{wiese2022identification,isit21,labidi2025joint,globecom_with_yaning,itw_with_yaning}.

\textbf{Control-Oriented Interpretation:}  
In the JDAI framework, identification provides a mechanism for subset selection. Each device performs a local binary decision (“Is this message intended for me?”), which directly maps to a control action. This interpretation is essential for scalable control under communication constraints.

\textbf{Practical Implementations:}  
Although identification theory is asymptotic, several practical constructions have been proposed, including structured codes such as Reed--Solomon and Reed--Muller codes, as well as combinatorial designs and pseudo-random implementations \cite{ferrara2022implementation,vonlengerke2023digital,vonlengerke2025tutorial,perotti2024wakeup,verdu1993explicit}. These approaches demonstrate that message identification paradigms can be realized under finite blocklength, hardware, and latency constraints.

\medskip

Overall, these properties indicate that message identification is particularly well suited for scenarios involving massive numbers of devices and stringent communication constraints, where classical transmission-based approaches become inefficient or infeasible.

\section{Joint Detection and Identification Framework}

In this section, we present the proposed \emph{Joint Detection and Identification} (JDAI) framework. The objective is to enable scalable monitoring and control of large populations of autonomous devices under severe communication constraints by tightly integrating sensing with message identification.

In contrast to classical communication architectures, which rely on explicit addressing and reliable message transmission, the JDAI framework adopts a fundamentally different paradigm: sensing is used to determine \emph{where} actions are required, while message identification determines \emph{which devices} should react. This decoupling is key to enabling scalable control in systems with massive numbers of devices.

\subsection{System Overview}

We consider a system consisting of a large population of autonomous devices (e.g., sensors, robots, or nanomachines) operating in a shared environment, referred to as the \emph{global space}. Within this space, we distinguish local \emph{regions of interest} in which specific actions should be triggered.

An external control system interacts with the device population via two complementary mechanisms:

\begin{itemize}
    \item \textbf{Joint detection (sensing):} The control system observes the global state and identifies regions in which relevant devices are present.
    \item \textbf{Message identification:} The control system broadcasts short signals that enable devices to locally decide whether to react.
\end{itemize}

\begin{figure}[t]
\centering
\includegraphics[width=0.9\textwidth]{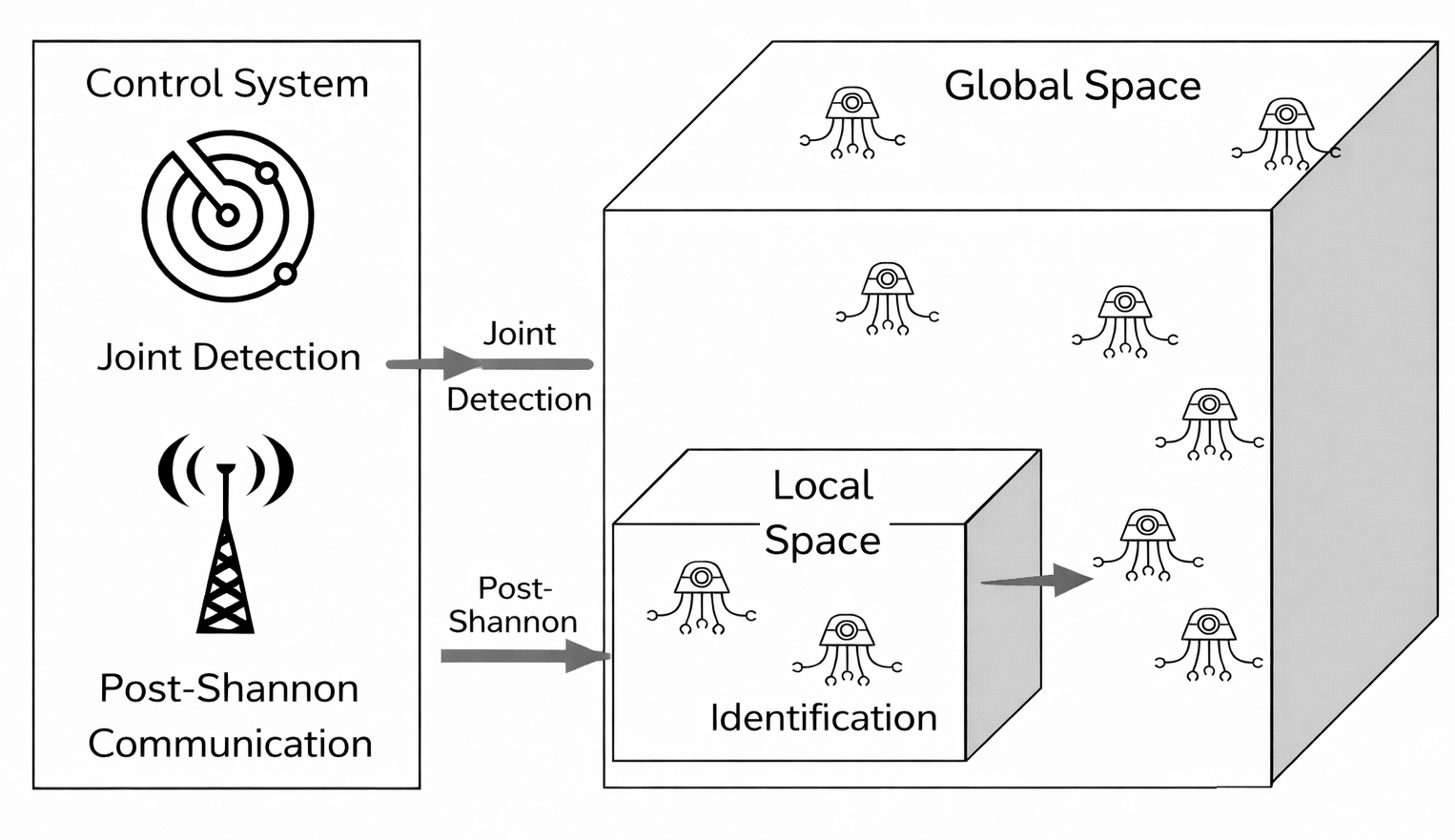}
\caption{Illustration of the JDAI framework. A large population of autonomous devices operates in a global space. The control system performs joint detection to identify regions of interest and broadcasts identification-based control signals. Although all devices receive the same signal, only a subset—determined by the identification mechanism—executes the corresponding action.}
\label{fig:JDAI}
\end{figure}

The overall architecture is illustrated in Fig.~\ref{fig:JDAI}. The key principle is that sensing and communication are tightly coupled: sensing determines when and where communication is required, while identification determines which devices should react.

\subsection{Autonomous Devices}

The system comprises a very large number of devices that are designed to perform simple, specialized tasks. Due to strict constraints on size, energy consumption, and hardware complexity, these devices typically have only limited capabilities \cite{akyildiz2008nanonetworks,akyildiz2010internet}.

In particular, devices may:
\begin{itemize}
    \item have only low-rate reception capabilities,
    \item lack continuous bidirectional communication,
    \item possess limited computational resources,
    \item operate passively (e.g., transported by environmental dynamics).
\end{itemize}

Each device is equipped with a receiver that allows it to process broadcast signals and perform identification-based decoding. Based on this decoding, the device decides whether a received message is relevant and whether a corresponding action should be executed.

Importantly, the number of devices is assumed to be sufficiently large such that classical addressing and individual control become infeasible.

\subsection{Joint Detection via Sensing}

The control system is assumed to have access to sensing mechanisms that provide information about the global state of the system. Depending on the application, these mechanisms may include electromagnetic, acoustic, optical, or biomedical sensing technologies \cite{farsad2016survey,nakano2013molecular,guo2018ultrasonic,sun2010magnetic,gleich2005tomographic}.

The purpose of sensing is not to track each individual device precisely, but to detect the presence of devices in regions of interest. This constitutes the \emph{joint detection} component of the framework, as it operates on the entire population rather than on individual devices.

More precisely, sensing provides a mapping from the global state to a set of candidate regions in which devices are likely to be present. This information allows the control system to determine when it is beneficial to initiate communication.

In many practical systems, sensing and communication are tightly coupled. For example, feedback signals or environmental observations can be used to generate common randomness, which in turn improves identification performance \cite{ezzine2020common,wiese2022identification,labidi2025joint,isit21,globecom_with_yaning,itw_with_yaning}.

\subsection{Message Identification}

Once a relevant region has been detected, the control system broadcasts a control signal encoded using identification codes \cite{Feedbackidentification,verdu1993explicit,Spandri2023,HashFctforIDcodes,OnurConstr}. Each device performs a local decision of the form:
\begin{center}
``Is this message intended for me?''
\end{center}

Based on this decision, the device either executes a predefined action or remains inactive.

A key advantage of this approach is that communication does not require explicit addressing of individual devices. Instead, identification codes allow the control system to implicitly select subsets of devices using short broadcast messages, even when the number of devices is extremely large.

The interpretation of identification messages is inherently probabilistic, giving rise to two types of errors:
\begin{itemize}
    \item \textbf{Missed activation} (error of the first kind),
    \item \textbf{False activation} (error of the second kind).
\end{itemize}

System design must ensure that both error probabilities remain within acceptable limits for the given application.

\subsection{Execution of Actions}

If a device identifies a received message as relevant, it executes a corresponding action. The nature of this action depends on the application and may include sensing, actuation, data collection, or release of a substance.

Importantly, successful system operation does not require that all intended devices react. In many scenarios, it is sufficient that at least one or a subset of devices within a region performs the desired action. This inherent redundancy significantly increases system robustness.

\subsection{Design Principles and Implications}

The JDAI framework induces several key design principles:

\textbf{Separation of Sensing and Communication:}  
Sensing identifies where actions are needed, while communication selects which devices should act.

\textbf{Broadcast-Based Control:}  
All devices receive the same signal, but only a subset reacts based on identification.

\textbf{Scalability:}  
Due to the properties of identification coding, the communication overhead grows only weakly with the number of devices.

\textbf{Robustness through Redundancy:}  
System reliability is achieved through the collective behavior of many devices rather than precise control of individual units.

\textbf{Technology-Agnostic Integration:}  
The framework is compatible with a wide range of physical-layer communication paradigms, including molecular, electromagnetic, acoustic, and magnetic communication systems \cite{farsad2016survey,nakano2013molecular,akyildiz2014internet,guo2018ultrasonic,sun2010magnetic}.

\medskip

Overall, the JDAI framework provides a system-level framework for controlling large populations of autonomous devices under severe communication constraints. It complements existing communication paradigms by introducing a control-oriented perspective based on identification.

\section{Biomedical Application: Targeted Nanorobot Therapy}

To illustrate the potential of the proposed JDAI framework, we consider a biomedical application in which a large population of micro- or nanoscale devices is deployed within the human body for targeted drug delivery. 
\begin{figure}[t]
    \centering
    \includegraphics[width=\linewidth]{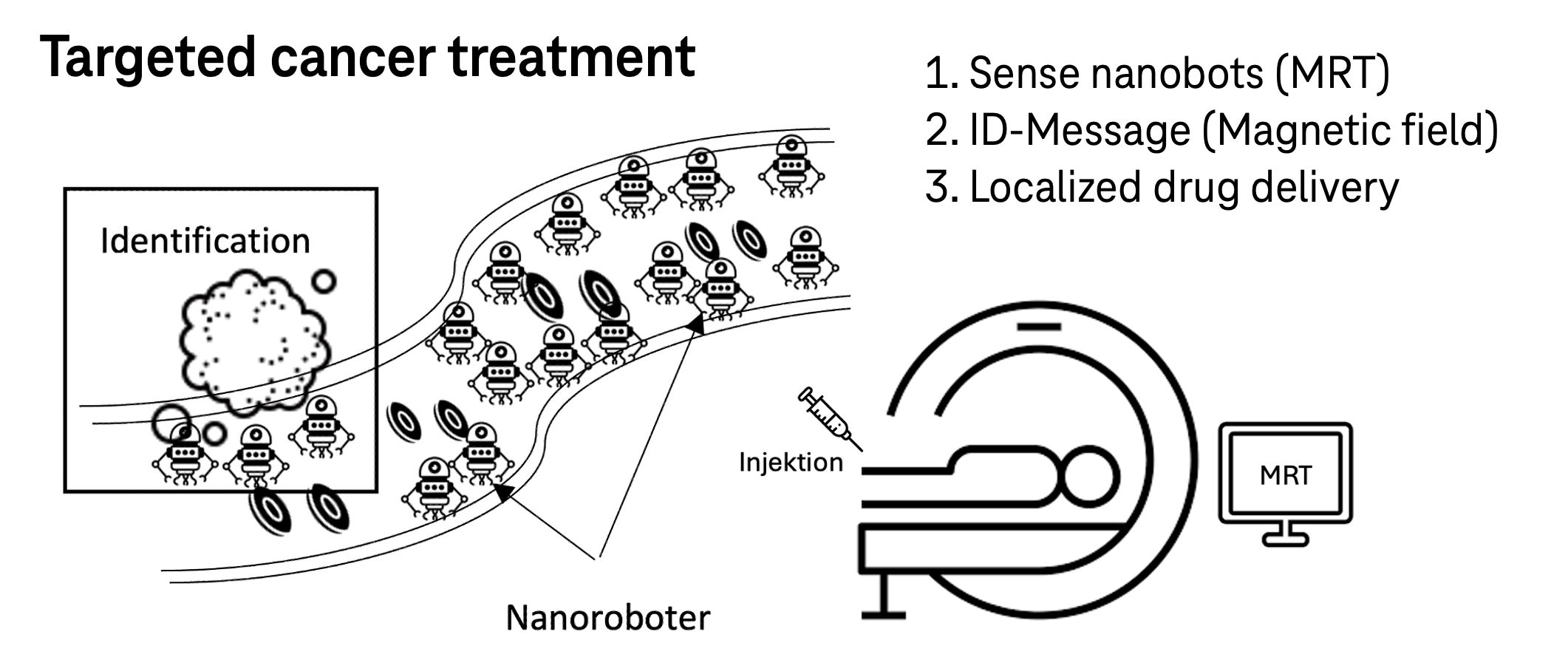}
    \caption{Illustration of targeted nanorobot therapy using the JDAI framework. 
    Nanorobots are injected into the bloodstream and transported through the vascular system. 
    An external control system performs sensing (e.g., MRI) to detect regions of interest (e.g., tumor sites) 
    and broadcasts identification-based signals to selectively activate a subset of devices.}
    \label{fig:nanorobot_scenario}
\end{figure}
This scenario highlights how the combination of sensing and message identification enables precise and scalable control under extremely constrained communication conditions.

In the considered setting, a large number of nanorobots is injected into the bloodstream of a patient and transported through the vascular system, which forms the \emph{global space}. A pathological region of interest, such as a tumor, defines a localized target region in which a therapeutic action should be triggered. To ensure that a sufficient number of devices reaches this region with high probability, a large population—potentially on the order of $10^5$ devices or more—is deployed. Due to strict constraints on size, energy, and hardware complexity, these nanorobots typically do not actively navigate but are instead transported passively by the blood flow \cite{sitti2009miniature,martel2012magnetic}.

An external control system, for example based on magnetic resonance imaging (MRI), performs both sensing and communication.\\ In a first step, the system estimates the spatial distribution of nanorobots within the body. Conventional MRI cannot resolve the nanorobots, as its resolution is limited by the wavelength, which is far too large for nanometer-scale machines. Therefore, a Small-Scale Magneto-Oscillatory (SMOL) device—a Radio Frequency (RF) sender integrated into the robot can be used. One drawback of this solution is the size of the RF sender, which is typically in the millimeter range, exceeding practical limits. However, MRI, does not detect objects based on size alone, but rather through the RF response of spins in a static gradient magnetic field. The external RF is used to excite the spins, creating an RF response signal, detectable by MRI. Different tissues produce different RF responses, resulting in different measurements. Spatial resolution is significantly improved, as it is defined by the gradient magnetic field and the RF resonance of body molecules within a longitudinal slice. Based on this principle, an alternative approach is to design the nanorobots such that they exhibit tissue-like or otherwise distinguishable resonance properties. By arranging the gradient magnetic field appropriately, the robots are detectable only when they come to the resonant zone (within a specific zone near the tumor). They would remain invisible most of the time, but when entering the resonant zone, they would emit RF signals detectable by MRI. This approach would allow selective detection of the nanorobots without requiring integrated RF senders.\\
In a second step, it transmits identification-based control signals that selectively activate those devices located in the region of interest. This two-stage process directly reflects the JDAI principle of separating detection and control while keeping them tightly coupled.

Each nanorobot can be modeled as a minimal autonomous unit with highly constrained resources. It comprises a receiving structure, for instance based on magnetic induction, that enables the detection of externally generated signals. A lightweight processing unit performs identification-based decoding, supported by a small memory module that stores pre-shared randomness. In addition, the device includes an actuation mechanism, such as a controlled drug release system, and operates under a severely limited energy budget, potentially supported by energy harvesting mechanisms.

Communication is realized via low-frequency or quasi-static magnetic fields generated by the external control system. This design choice is motivated by the strong attenuation of high-frequency electromagnetic signals in biological tissue, which makes conventional wireless communication infeasible in such environments \cite{sun2010magnetic}. As a result, the achievable communication rates are extremely limited, which further motivates the use of identification-based signaling rather than classical transmission-based approaches.

The sensing functionality is integrated into the external system, for example through MRI-based techniques \cite{gleich2005tomographic}. The sensing process excites the nanorobots using an external magnetic field, after which the devices exhibit characteristic responses that can be measured and processed to infer their spatial distribution. By exploiting differences in resonance behavior or induced signals, it becomes possible to identify regions in which nanorobots are present with sufficiently high probability. Importantly, the objective is not to track each individual device precisely, but rather to detect the presence of devices in regions of interest. This corresponds directly to the joint detection component of the JDAI framework. The achievable spatial resolution depends on factors such as signal-to-noise ratio, acquisition time, and the physical properties of the sensing modality, which introduces a fundamental trade-off between sensing accuracy and latency.

A key challenge in the considered system is the sensing problem itself. 
Due to fundamental physical limitations, individual nanorobots cannot be directly resolved by imaging modalities such as MRI. 
Instead, detection relies on indirect mechanisms, such as collective effects or engineered resonance properties of the devices. 
As a result, sensing provides only coarse and probabilistic information about the spatial distribution of the devices. 
This uncertainty directly impacts the performance of the JDAI framework, as it affects both the identification stage and the reliability of control decisions.

Once the control system detects that nanorobots are present in the target region, it initiates communication by broadcasting an identification message. The communication is implemented via magnetic field modulation, where the induced voltage at a nanorobot can be approximated as
\begin{equation}
U = A \cdot \frac{dB}{dt},
\end{equation}
with $A$ denoting the effective receiving area. Each nanorobot evaluates the received signal using its locally stored randomness and performs a binary identification test. If the message is relevant, the device activates its drug release mechanism; otherwise, it remains inactive.

A central advantage of message identification lies in its favorable scaling behavior. While classical transmission schemes require communication resources that grow logarithmically with the number of possible messages, identification allows a double-logarithmic scaling. For example, for a system with $10^6$ possible identities, the corresponding identification rate scales as $\log \log (10^6) \approx 1.7$ bits. This value should be interpreted as an information-theoretic indicator rather than a direct finite-blocklength requirement, since practical implementations must account for noise, reliability constraints, and coding overhead. Nevertheless, it illustrates that even extremely low-rate communication—such as magnetic field modulation at a few Hertz—can enable timely control actions.

The overall operational workflow can be described as a closed-loop process in which sensing and communication are tightly integrated. After deployment of the nanorobots, the control system continuously monitors their spatial distribution, detects regions of interest, and subsequently triggers identification-based control signals. The activated nanorobots then execute the desired action, such as releasing a therapeutic agent. This  reflects the key idea of JDAI; sensing determines when and where to act, while identification determines which devices should act.

The presented application highlights several important aspects of the JDAI framework. First, it demonstrates that message identification enables scalable control of extremely large device populations under severe communication constraints. Second, it shows that system robustness can be achieved through redundancy, since successful operation does not depend on the precise behavior of individual devices but rather on the collective response of many agents. Third, it indicates that the framework is compatible with current technological developments, including MRI-based sensing, magnetic field communication, and advances in nanotechnology \cite{sitti2009miniature,martel2012magnetic}.

At the same time, the example remains intentionally simplified and points to several important extensions. In realistic scenarios, the state of the system is typically continuous-valued, for example in terms of spatial position or velocity, and the communication channel is better modeled by Gaussian or fading channel models. Moreover, interactions between multiple devices may lead to interference effects that require multiuser communication models. Finally, practical implementations must account for finite blocklength constraints and latency requirements, which play a crucial role in real-time biomedical applications. These aspects motivate future work on extending the JDAI framework to more realistic communication and sensing models while preserving its scalability and control-oriented structure.

\section{Feasibility and Technical Considerations}

In this section, we discuss the feasibility of the proposed JDAI framework in light of current technological capabilities. While a complete end-to-end realization of such a system is not yet available, many of its individual components have already been demonstrated in isolation. The main challenge therefore lies not in the absence of enabling technologies, but in their integration into a coherent and highly constrained system architecture.

From an information-theoretic perspective, identification coding has been extensively studied and is well understood in terms of its asymptotic performance limits \cite{ahlswede1989identification,han2003information}. More recently, practical constructions based on structured codes and pseudo-random implementations have been developed, demonstrating that message identification can be realized under realistic constraints such as finite blocklength, limited computational resources, and hardware imperfections \cite{ferrara2022implementation,vonlengerke2023digital,vonlengerke2025tutorial,perotti2024wakeup,verdu1993explicit}. These developments indicate that identification coding is not merely of theoretical interest, but can serve as a practical building block in real-world systems.

On the sensing side, technologies capable of detecting weak signals with high spatial resolution are already widely available. In particular, magnetic resonance imaging (MRI) systems are able to generate controlled magnetic fields and measure very small magnetic responses with high sensitivity \cite{gleich2005tomographic}. Such systems naturally provide both the sensing and actuation capabilities required for the JDAI framework. In addition, other sensing modalities, including acoustic, optical, and biochemical methods, have been studied extensively in the context of nanoscale and in-body communication systems \cite{farsad2016survey,nakano2013molecular,guo2018ultrasonic,akyildiz2014internet}. This suggests that the sensing component of JDAI can be implemented using a variety of physical platforms, depending on the application.

At the level of nanoscale devices, significant progress has been made in micro- and nanotechnology, including the development of micro-scale sensors, simple processing units, and energy harvesting mechanisms \cite{sitti2009miniature,martel2012magnetic}. Individual components such as signal receivers, basic logic circuits, and actuation mechanisms have already been demonstrated. However, the integration of all these components into a single nanoscale device remains a major engineering challenge. In particular, constraints on size, energy consumption, and biocompatibility impose strict limitations on system design.

Energy availability constitutes one of the most critical constraints. Nanorobots must operate under extremely limited energy budgets, which restricts both sensing and communication capabilities. This limitation reinforces the relevance of message identification, since it minimizes the required communication overhead while still enabling effective control. At the same time, it necessitates highly efficient implementations of decoding algorithms and possibly the use of energy harvesting techniques.

Another important aspect concerns the reliability of identification-based decisions. By design, identification allows for probabilistic errors of the first and second kind. In safety-critical applications such as targeted drug delivery, these errors must be carefully controlled. However, the JDAI framework inherently provides robustness through redundancy, since successful operation does not rely on the correct behavior of individual devices but rather on the collective response of many devices within a region of interest. This system-level redundancy can be exploited to mitigate the impact of individual decoding errors.

From a system perspective, several additional challenges arise. The interaction between sensing and communication introduces timing constraints, since delays in sensing or actuation may affect system performance. For example, the movement of nanorobots due to blood flow implies that the spatial distribution of devices may change between sensing and actuation, which must be taken into account when designing control strategies. Furthermore, synchronization between sensing and communication phases is required to ensure consistent system behavior.

Another key challenge is scalability. While identification coding provides favorable scaling properties from an information-theoretic perspective, practical implementations must ensure that sensing, signal processing, and control mechanisms can handle large populations of devices in real time. This includes efficient processing of sensing data and the design of identification signals that remain robust under realistic noise and interference conditions.

Biocompatibility and safety considerations are also of central importance in biomedical applications. All components of the system, including nanorobots and externally applied fields, must satisfy strict safety constraints. In particular, the strength and frequency of electromagnetic or magnetic fields must remain within medically acceptable limits, and the materials used for nanorobots must be biocompatible.

Finally, it is important to emphasize that the JDAI framework does not depend on a specific technology and can be extended beyond the specific biomedical scenario considered in this paper. The same principles apply to other large-scale systems, including Internet-of-Things (IoT) networks, distributed sensor systems, and autonomous robotic swarms. In all these scenarios, the combination of sensing and message identification provides a promising approach to scalable control under communication constraints.

Overall, the JDAI framework is feasible since its key components—identification coding, sensing technologies, and nanoscale devices—are already available or under active development. The main open challenge lies in their integration into a unified system that satisfies the strict constraints of real-world applications. From both an engineering and an information-theoretic perspective, the presented results indicate that such integration is challenging but achievable, and that message identification can play a central role in enabling future large-scale control systems.

\section{Conclusion}

In this paper, we introduced the concept of Joint Detection and Identification (JDAI) as a framework for the scalable monitoring and control of large populations of autonomous devices under severe communication constraints. The key idea is to combine sensing-based detection with message identification, thereby replacing classical individual addressing by a broadcast paradigm in which only selected subsets of devices react.

The theory of identification \cite{ahlswede1989identification,han2003information}, differs fundamentally  from classical transmission. In particular, the double-exponential scaling of identification messages enables efficient selection among extremely large sets of possible actions, even when the underlying communication channel is highly constrained. This makes identification particularly attractive for emerging applications with limited resources and a large number of devices.

From a system perspective, we showed that identification can be interpreted as a mechanism for subset selection, where devices locally decide whether a broadcast signal is relevant. This interpretation allows the integration of identification into control-oriented architectures and provides a natural bridge between information-theoretic principles and practical system design.

Based on these insights, we proposed the JDAI framework, which decouples the control process into two tightly coupled components: a sensing stage that identifies regions of interest, and a communication stage that enables devices to locally decide whether to act. This architecture avoids the need for explicit addressing and enables scalable control in large populations of simple and resource-constrained devices.

We illustrated the proposed framework using a biomedical application scenario involving nanorobots for targeted drug delivery. In this setting, sensing mechanisms such as MRI provide coarse spatial information, while message identification enables selective activation of devices under extremely limited communication rates. The example demonstrates how the JDAI principle can be realized in a physically constrained environment and highlights its potential for practical applications.

Furthermore, we discussed feasibility aspects and showed that the key components of the framework—identification coding, sensing technologies, and nanoscale devices—are already available or under active development. While significant challenges remain, particularly in terms of system integration, energy constraints, and reliability, the presented results indicate that JDAI provides a promising and realistic approach to scalable control in future systems.

Several directions for future work arise naturally from this study. On the theoretical side, extending the framework to continuous-valued state spaces and Gaussian channel models is of particular interest. In addition, multiuser settings and interference-limited scenarios require further investigation to understand the impact of interactions between devices. From a practical perspective, finite blocklength constraints, latency requirements, and hardware limitations must be incorporated into the design of message identification schemes. Finally, the integration of sensing, communication, and actuation into unified system architectures remains a central challenge that calls for interdisciplinary research.

Overall, the JDAI framework highlights a shift in perspective: rather than transmitting detailed information to each device, control is done through global observation and minimal, identification-based signaling. This paradigm is particularly well suited for emerging large-scale systems, including biomedical applications, Internet-of-Things networks, and distributed autonomous systems. This opens new directions for the design of communication and control architectures under extreme constraints.

\section*{Acknowledgment}

The authors gratefully acknowledge the financial support of the Federal Ministry of Research, Technology and Space of Germany (BMFTR) within the programme “Souverän. Digital. Vernetzt.”, joint project 6G-life (grant numbers 16KIS2414 and 16KIS2415).
This work was further supported by the BMFTR Quantum Programme, including the projects QUIET (grants 16KISQ093 and 16KISQ0170), QD-CamNetz (grants 16KISQ077 and 16KISQ169), and QSTARS (grants 16KIS2611 and 16KIS2602).
Additional support was provided by the German Research Foundation (DFG) within the project “Post-Shannon Theory and Implementation” (grants DE1915/2-1 and BO 1734/38-1).
The authors also acknowledge financial support from the Federal Ministry of Education and Research of Germany (BMBF) within the project Internet of Bio-Nano-Things (IoBNT) under grant number 5310223.
% Use IEEE DIN 1505 style for bibliography / Literaturverzeichnisses
        \bibliographystyle{unsrt}
        %\nocite{*}              % Include all references without checking / Alle References immer aufführen
        \bibliography{references}

\end{document}